# Enhanced photoluminescence from single nitrogen-vacancy defects in nanodiamonds coated with metal-phenolic networks

*Kerem Bray[1], Rodolfo Previdi[1], Brant C. Gibson[2], Olga Shimoni[1]\*, and Igor Aharonovich[1]\**

[1] School of Physics and Advanced Materials, University of Technology, Sydney, P.O. Box 123, Broadway, New South Wales 2007, Australia

[2] ARC Centre of Excellence for Nanoscale BioPhotonics, School of Applied Sciences, RMIT University, Melbourne, VIC 3001, Australia

\* Olga.Shimoni@uts.edu.au , Igor.Aharonovich@uts.edu.au

ABSTRACT

Fluorescent nanodiamonds are attracting major attention in the field of bio-sensing and bio-labeling. In this work we demonstrate a robust approach to surface functionalize individual nanodiamonds with metal-phenolic networks that enhance the photoluminescence from single nitrogen vacancy (NV) centers. We show that single NV centres in the coated nanodiamonds also exhibit shorter lifetimes, opening another channel for high resolution sensing. We propose that the nanodiamond encapsulation suppresses the non-radiative decay pathways of the NV color centers. Our results provide a versatile and assessable way to enhance photoluminescence from nanodiamond defects that can be used in a variety of sensing and imaging applications.

Over the last few decades, rapid advances in the life sciences have driven the demand for the development of highly sensitive probes for bio-imaging and sensing. Currently, fluorescence microscopy is the most abundant technique to get an insight into molecular dynamics with probes spanning from fluorescent organic molecules to nanoparticles, including fluorescent proteins or organic molecules, quantum dots, upconversion nanocrystals and others.[1-3] Despite numerous advantages for each system, most of them exhibit undesirable properties, such as fast photobleaching, blinking or toxicity, which diminishes their use in practical applications. Recently, nanodiamond particles have emerged as a promising nanomaterial for bio-imaging and bio-sensing as they possess a number of benefits including inherent biocompatibility, carbon surface for ease of biofunctionalization and photostability.[4-6] In addition, they can host photostable optically active defects, such as the nitrogen vacancy (NV) or the silicon vacancy (SiV) that have already been shown as effective bioimaging probes.[7-10] Moreover, single NV defects can be leveraged for high resolution magnetometry and sensing of single spins and individual biomolecules with unprecedented sensitivity.[10-14] Nevertheless, the optical detection of single fluorescent NV centers is challenging with a standard confocal microscope, due to the relatively long fluorescence lifetime (~ 11 – 22 ns) that reduces the overall flux of photons comparing with standard florescent molecular probes (1 – 5 ns).[2, 15] Therefore, to establish reliable and efficient probes based on the single NV center for magnetometry and bio-imaging, there is a demand to enhance the NV luminescence.

So far, most of the techniques to achieve enhancement in NV photoluminescence in nanodiamonds are based on near field effects or high energy electron irradiation. In first case, the nanodiamond is brought in a close proximity to a metallic structure to achieve plasmonic coupling,[16] or positioned on top of a dielectric cavity.[17] While these techniques are suitable for nanophotonics and quantum optics, they are performed on an extremely limited number of particles at a time and are not appropriate for bio-sensing or bio-imaging. In the latter case, high energy electron beam irradiation creates ensembles of NV centers, and the important attributes of single photon emission and high contrast in the optically detected magnetic resonance (ODMR) signal are lost.

In this work, we demonstrate that the photoluminescence of a single NV defect embedded in a 45 nm nanodiamond can be significantly enhanced using a simple, cost effective coating method with a metalo-phenolic network. In addition, we show that the photoluminescence enhancement is accompanied by a fluorescence lifetime reduction and an excellent signal-to-noise ratio in the

ODMR signal from a single NV center. These attributes are important for utilizing the signal from the NV center for sensing applications. Furthermore, our method enables biocompatible and functional coating of bulk quantities of single nanodiamonds and therefore can be implemented to improve nanodiamond use for bio-imaging, bio-sensing and magnetometry.

The coating approach is based on the recently established method that is a robust one-step self-assembly between metal ions and phenol compound (tannic acid, TA).[18, 19] Figure 1a shows the schematic presentation of the process. To produce the metalo-phenolic coated nanodiamonds, 500 µL of the nanodiamond (NaBond) solution (0.5 mg/ml in MilliQ water) were individually mixed with 5 µL of 10 mg/mL $FeCl_3 \cdot 6H_2O$ (Fe (III)) and 5 µL of the 40 mg/mL TA solutions upon vigorous agitation. This addition caused an almost instantaneous color change from clear to purple-grey. In the final step, 500 µL of a 3-(N-morpholino)propanesulfonic acid buffer was added to the solution to ensure formation of stable networks and adjusted PH to 7.4. The sample then underwent two washing/centrifugation cycles with water to remove excess of iron (III) and TA (10 000 rcf for 10 min). Formation of the metal-phenolic complex on the nanodiamonds was monitored using zeta potential that indicated a change from –13.5±4.8 mV to –25.4±6.4 mV due to a number of galloyl groups on TA. As the starting materials are biocompatible (nanodiamonds have been shown to be non-toxic,[20] and components of the metalo-phenolic network are recognized safe by the U.S. Food and Drug Administration) the formed complex has been shown to exhibit biocompatibility property as well,[19] meaning it is safe to use in biological applications.

To confirm the successful surface modification of nanodiamonds, we performed energy-dispersive X-ray spectroscopy (EDS) mapping using scanning electron microscopy (SEM, Zeiss Evo LS15) equipped with Bruker xFlash detector. The measurements were performed at room temperature with 5 kV as accelerating voltage. For ease observation of element maps and due to the limited resolution of the system, the metal-organic complexes were assembled on 500 nm diamond particles that were dispersed on silicon substrates. SEM images of coated and pristine nanodiamonds can be seen in Fig. 1b (i) and (iv). EDS mapping analysis of both pristine and coated nanodiamond for carbon (Fig. 1b (ii) and (v), red) exhibited a good alignment with SEM images of the same area. EDS mapping analysis for oxygen revealed that there are matching oxygen patterns for the nanodiamonds with metal-organic complexes (Fig. 1b (iii), green), while uncoated nanodiamonds has a negligible detected amount of oxygen (Fig. 1 b (vi)). The increased amount of oxygen on the coated nanodiamonds is due to multiple oxygen atoms in the

TA molecule (46 atoms of oxygen atoms *versus* 76 of carbon atoms) leading to higher detection level. High amount of oxygen at the background is the result of native silicon oxide on the silicon substrate (silicon from the substrate was detected as well, not shown).

Fig. 1b shows the NV photoluminescence (PL) spectrum from a 45 nm nanodiamond conjugated with the Fe(III)-TA complex, recorded at room temperature using a 532 nm excitation wavelength. The zero phonon line (ZPL) of the negative charged NV center is visible at 637 nm, and the spectral properties are not hindered by the presence of the complex. The complex itself does not fluoresce at the same spectral window.

To provide a quantitative analysis of the complex's influence on the emission properties of NV defects, only nanodiamonds that host single NV emitters were investigated. For the optical measurements, a home built scanning confocal microscope equipped with a Hunbury Brown & Twiss interferometer with avalanche photo diodes (APD, SPCM-AQRH-14) for single photon detection and a spectrometer (Princeton Instruments) was utilized. All the measurements were carried out at room temperature using a 532 nm excitation through a high numerical aperture (NA=0.9) objective. A fiber splitter was used to direct the signal either to the spectrometer or to the APDs. The schematics of the setup is shown in Fig. 2a.

Fig. 2b,c shows a normalized second order auto-correlation function, $g^{(2)}(\tau)=<I(t)I(t+\tau)>/<I(t)>^2$, recorded from a typical pristine nanodiamond and from a typical nanodiamond coated with the metal-organic complexes, respectively. The dip below 0.5 at zero delay time *(t=0)* indicates that the signal is collected from a single NV center. The data is fit using a standard three level system $g^{(2)}(\tau)=1-(1+a)exp(-\lambda_1\tau)+aexp(-\lambda_2\tau)$, where $\lambda_1$, $\lambda_2$ are related to the decay times from the excited to the metastable states and from the excited state to the ground state.

One of the important characteristics of NV centers is its utilization for magnetic sensing due to its spin dependent fluorescence. Specifically, under zero magnetic field, the NV ground state exhibits splitting of 2.87 GHz (Fig. 3a) between the $m_s = 0$ and $m_s = \pm 1$ spin sub-levels of the spin triplet ground state.[21] The spin preserving transitions from the excited state enable the spin state to be read at room temperature employing an optically detected magnetic resonance (ODMR) technique.

To perform the ODMR measurements, a microwave pulse generator (Rhodes & Shwartz SM300) was employed. Fig. 3b,c shows the typical ODMR measurement recorded from a single NV center embedded in a pristine and conjugated nanodiamond, respectively. The fluorescence contrast at a resonance frequency of 2.87 GHz is clearly observed and can be described using Lorentzians fitting. The splitting is most likely originated due to residual strain fields within the nanodiamonds or stray magnetic fields. The importance of this measurement lies with the fact that the high visibility of the ODMR signal is maintained despite the conjugation of the nanodiamonds.

To examine the influence of the metal-organic complex on the NV center lifetime, time resolved measurements were performed. Fig. 4a shows the lifetime recorded from a typical pristine (red circles) and a coated nanodiamond (black squares) hosting single NV centers, respectively. The measurements are performed at room temperature using a picosecond laser diode ($\lambda$=514 nm). The data can be fitted well using a double exponential fitting (solid lines). Fig. 4 (b, c) shows the histogram of the fluorescence lifetimes from pristine and conjugated nanodiamonds hosting single emitters, respectively. The average decay time of the coated nanodiamonds is faster with a lifetime of $\tau$=5.45±1.49 ns, while the pristine nanodiamonds exhibit a lifetime of $\tau$=8.31±1.05 ns. Faster lifetime is often associated with brighter fluorescence since more photons are emitted, as indeed observed from the nanodiamonds with the metal-organic networks.

Fig. 4d shows an intensity time trace recorded from a typical pristine (red curve) and a typical conjugated nanodiamond (black curve). It can be clearly observed that the complex does not induce bleaching or blinking and the investigated nanodiamonds are photostable, confirming their superior behavior as a probe for bio-labeling. Moreover, the coated nanodiamonds exhibit at least a threefold increase in their fluorescence compared with the pristine nanodiamonds. As both measurements were performed on nanodiamonds with single emitters (with and without complex), using same excitation power, this indicates that the complex directly enhances the fluorescence from the nanodiamonds. We studied more than 25 conjugated nanodiamonds with single emitters and all showed brighter fluorescence, under the same excitation conditions compared to pristine nanodiamonds. We note that due to the unknown emission dipole within the nanodiamonds, the intensity enhancement can vary.

To understand the underlying process for the observed photoluminescence enhancement from the coated nanodiamond, we further discuss the potential mechanisms. It is unlikely that Fluorescence Resonance Energy Transfer (FRET) process occurs since the metal-organic complex does not emit itself under the green laser illumination, and it has no spectral overlap with the NV center emission. Additionally, absorption band of the complexes does not overlap with the NV spectrum – as requirement for an efficient FRET[22]. Moreover, we have not observed any intensity variations (i.e. blinking) as typically observed for FRET processes.

A more plausible explanation for the enhanced brightness is the suppression of non-radiative recombination sites,[23] that often present on the nanodiamond surfaces. The total decay rate from a single emitter can be described by $1/\tau = 1/\tau_{rad} + 1/\tau_{nrad}$ where $\tau_{nrad}$ and $\tau_{rad}$ are the non-radiative and the radiative decay times, respectively, while $\tau$ is the total measured decay time. Since the immediate dielectric environment of the nanodiamonds was not modified,[23, 24] we assume that the radiative decay remains the same for the coated and the pristine nanodiamonds. Therefore, the non-radiative component must be modified since the overall decay time has changed. The nanodiamond surface often has charge traps that involve non-radiative recombination pathways.[25] Upon introduction of the coating of the molecule, such traps are suppressed and several non-radiative recombination channels are eliminated. Since we do not know the precise dipole orientation of every single emitter that is probed, we cannot directly calculate the radiative and the non-radiative components. However, our measurements indicate that the quantum efficiency of a single NV center in a nanodiamond is not unity, and can be improved significantly using a coating with organic molecules, resulting in brighter fluorescent nanodiamonds.

To conclude, we presented a robust, single step, cost effective process to enhance nanodiamonds fluorescence through the coating with metal-organic complexes. We show that the photostability, the ODMR signal and emission spectra of the nanodiamonds are not affected. Moreover, the photoluminescence from the single NV centers is enhanced due to suppression of non-radiative channels. Due to the low cost, ease of fabrication, scalability and negligible cytotoxicity, the described approach can be utilized as a platform for the assembly of advanced materials for potential use in a versatile range of bio-applications.

Fig.s

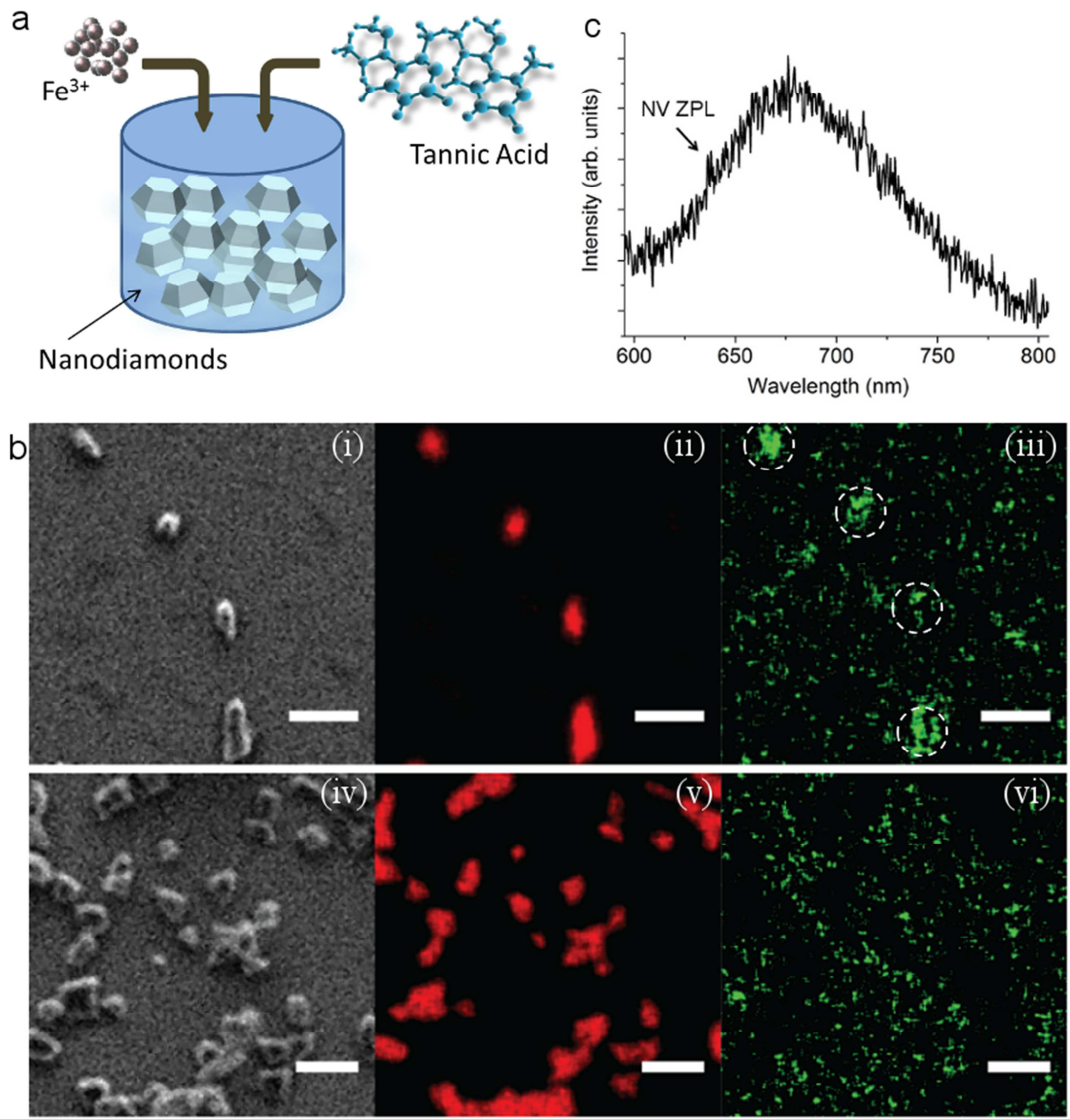

*Fig. 1*. *Complex preparation and photoluminescence spectrum. (a) Schematic illustration of the complex preparation and the coating process of the nanodiamonds. A one-step assembly of coordination complexes on a substrate through the mixing of TA, iron(III) and nanodiamonds with NV centers; (b) SEM and EDS elemental maps of coated ((i) – (iii)) and uncoated ((iv) – (vi) nanodiamonds with metalo-organic complexes. Red color ((ii) and (v)) corresponds to carbon content, while green color ((iii) and (vi)) corresponds to oxygen atoms. Scale bar is 2 μm (c) Room temperature photoluminescence spectrum recorded from a conjugated nanodiamond hosting a single NV center.*

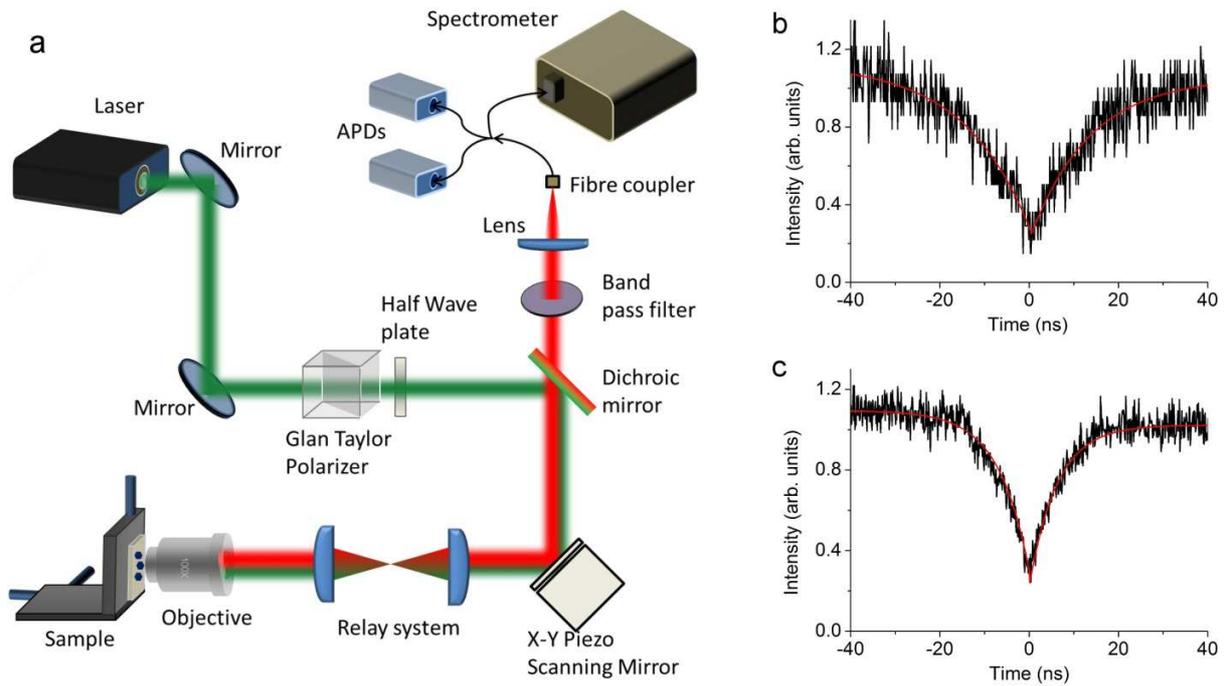

*Fig. 2*. *Optical properties of the nanodiamonds (a) Schematic illustration of the scanning confocal microscope setup. Green laser (continuous or pulsed) is used for excitation through a high numerical aperture objective (NC=0.9). The signal is coupled into an optical fiber that acts as a confocal aperture and directed into the APDs or a spectrometer. (b, c) Second order correlation function, $g^{(2)}(\tau)$, recorded from a pristine and coated nanodiamond hosting NV centers, respectively. The dip in $g^{(2)}(\tau)$ below 0.5 indicates that the addressed defects are single photon emitters.*

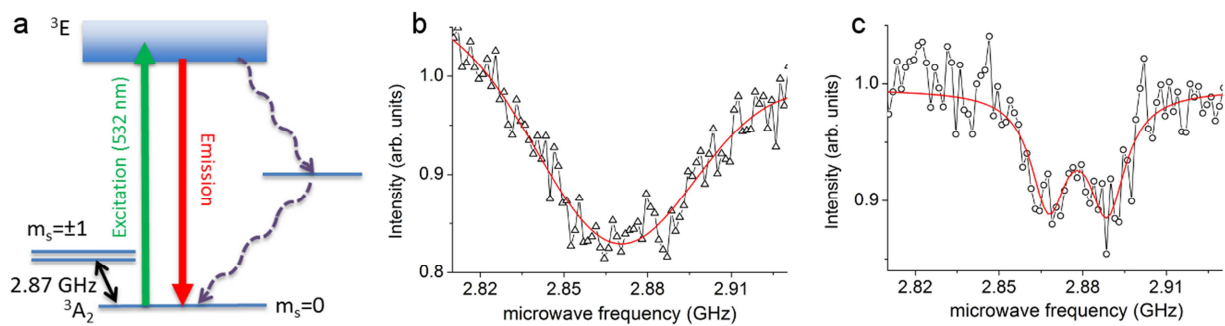

*Fig. 3*. *Optically detected magnetic resonance.* *(a) Energy level diagram of the NV center. At zero magnetic field, the ground state is split by 2.87 GHz that give rise to the ODMR signal. (b, c) ODMR measurements from a typical pristine and a typical coated nanodiamond hosting a single NV center, respectively.*

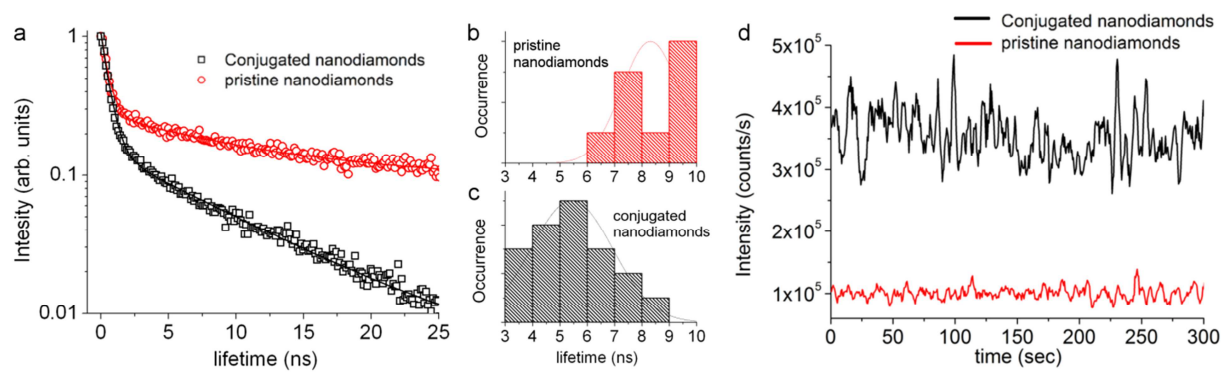

*Fig. 4*. *Fluorescence lifetime and intensity measurements. (a) Lifetime measurement of a typical pristine (red circles) and a typical coated nanodiamond (black squares) hosting single NV centers, respectively. The decay is faster from the conjugated nanodiamonds. The solid lines are fit to data. (b,c) lifetime histogram from pristine and conjugated nanodiamonds. More than 25 single emitters were studied for each group. (d) Intensity time trace confirming that a typical conjugated nanodiamond exhibit photostable and brighter photoluminescence compared to that measured for a typical pristine nanodiamond.*

**ACKNOWLEDGMENT**

Brant Gibson is the recipient of an Australian Research Council Future Fellowship (Project Number FT110100225). Olga Shimoni acknowledges the Ramaciotti foundation for the financial support. Igor Aharonovich is the recipient of an Australian Research Council Discovery Early Career Research Award (Project Number DE130100592). We thank Charlene Lobo and Russell Sandstorm for useful discussions.

REFERENCES

1. F. Wang, D. Banerjee, Y. Liu, X. Chen and X. Liu, *Analyst*, 2010, **135**, 1839-1854.
2. P. Alivisatos, *Nature Biotechnology*, 2004, **22**, 47-52.
3. Y. Lu, J. Zhao, R. Zhang, Y. Liu, D. Liu, E. M. Goldys, X. Yang, P. Xi, A. Sunna, J. Lu, Y. Shi, R. C. Leif, Y. Huo, J. Shen, J. A. Piper, J. P. Robinson and D. Jin, *Nature Photon.*, 2014, **8**, 32-36.
4. V. N. Mochalin, O. Shenderova, D. Ho and Y. Gogotsi, *Nature Nanotech.*, 2012, **7**, 11-23.
5. A. Krueger and D. Lang, *Adv. Funct. Mater.*, 2012, **22**, 890-906.
6. I. Aharonovich and E. Neu, *Advanced Optical Materials*, 2014, **2**, 911-928.
7. T. D. Merson, S. Castelletto, I. Aharonovich, A. Turbic, T. J. Kilpatrick and A. M. Turnley, *Optics Letters*, 2013, **38**, 4170-4173.
8. S. Kaufmann, D. A. Simpson, L. T. Hall, V. Perunicic, P. Senn, S. Steinert, L. P. McGuinness, B. C. Johnson, T. Ohshima, F. Caruso, J. Wrachtrup, R. E. Scholten, P. Mulvaney and L. Hollenberg, *Proc. Natl. Acad. Sci.*, 2013, **110**, 10894-10898.
9. N. Mohan, Y. K. Tzeng, L. Yang, Y. Y. Chen, Y. Y. Hui, C. Y. Fang and H. C. Chang, *Adv. Mater.*, 2009, **22**, 843.
10. A. Ermakova, G. Pramanik, J. M. Cai, G. Algara-Siller, U. Kaiser, T. Weil, Y. K. Tzeng, H. C. Chang, L. P. McGuinness, M. B. Plenio, B. Naydenov and F. Jelezko, *Nano Letters*, 2013, **13**, 3305.
11. V. R. Horowitz, B. J. Alemán, D. J. Christle, A. N. Cleland and D. D. Awschalom, *Proc. Natl. Acad. Sci.*, 2012, **109**, 13493–13497.
12. G. Balasubramanian, I. Y. Chan, R. Kolesov, M. Al-Hmoud, J. Tisler, C. Shin, C. Kim, A. Wojcik, P. R. Hemmer, A. Krueger, T. Hanke, A. Leitenstorfer, R. Bratschitsch, F. Jelezko and J. Wrachtrup, *Nature*, 2008, **455**, 648-652.
13. M. S. Grinolds, S. Hong, P. Maletinsky, L. Luan, M. D. Lukin, R. L. Walsworth and A. Yacoby, *Nature Phys.*, 2013, **9**, 215-219.
14. M. Geiselmann, M. L. Juan, J. Renger, J. M. Say, L. J. Brown, F. J. G. de Abajo, F. Koppens and R. Quidant, *Nature Nanotech.*, 2013, **8**, 175–179
15. O. Faklaris, D. Garrot, V. Joshi, J. P. Boudou, T. Sauvage, P. A. Curmi and F. Treussart, *Journal of the European Optical Society-Rapid Publications*, 2009, **4**, 09035.
16. M. Barth, S. Schietinger, S. Fischer, J. Becker, N. Nusse, T. Aichele, B. Lochel, C. Sonnichsen and O. Benson, *Nano Letters*, 2010, **10**, 891-895.

17. J. Wolters, A. Schell, G. Kewes, N. Nusse, M. Schoengen, H. Doscher, T. Hannappel, B. Lochel, M. Barth and O. Benson, *Appl. Phys. Lett.*, 2010, **97**, 141108.
18. J. Guo, Y. Ping, H. Ejima, K. Alt, M. Meissner, J. J. Richardson, Y. Yan, K. Peter, D. von Elverfeldt, C. E. Hagemeyer and F. Caruso, *Angewandte Chemie*, 2014, **126**, 5652-5657.
19. H. Ejima, J. J. Richardson, K. Liang, J. P. Best, M. P. van Koeverden, G. K. Such, J. Cui and F. Caruso, *Science*, 2013, **341**, 154-157.
20. C. C. Fu, H. Y. Lee, K. Chen, T. S. Lim, H. Y. Wu, P. K. Lin, P. K. Wei, P. H. Tsao, H. C. Chang and W. Fann, *Proceedings of the National Academy of Sciences of the United States of America*, 2007, **104**, 727-732.
21. M. W. Doherty, N. B. Manson, P. Delaney and L. C. L. Hollenberg, *New J. Phys.*, 2011, **13**, 025019.
22. J. Tisler, R. Reuter, A. Lämmle, F. Jelezko, G. Balasubramanian, P. R. Hemmer, F. Reinhard and J. Wrachtrup, *ACS Nano*, 2011, **5**, 7893-7898.
23. F. A. Inam, M. D. W. Grogan, M. Rollings, T. Gaebel, J. M. Say, C. Bradac, T. A. Birks, W. J. Wadsworth, S. Castelletto, J. R. Rabeau and M. J. Steel, *ACS Nano*, 2013, **7**, 3833-3843.
24. F. A. Inam, T. Gaebel, C. Bradac, L. Stewart, M. J. Withford, J. M. Dawes, J. R. Rabeau and M. J. Steel, *New J. Phys.*, 2011, **13**, 073012.
25. C. Bradac, T. Gaebel, C. I. Pakes, J. M. Say, A. V. Zvyagin and J. R. Rabeau, *Small*, 2012, **9**, 132–139.

**Supporting information for *"Enhanced photoluminescence from single nitrogen-vacancy defects in nanodiamonds coated with phenol-ionic complexes"***

*Kerem Bray[1], Rodolfo Previdi[1], Brant C. Gibson[2], Olga Shimoni[1]\*, and Igor Aharonovich[1]\**

[1] School of Physics and Advanced Materials, University of Technology, Sydney, P.O. Box 123, Broadway, New South Wales 2007, Australia

[2] ARC Centre of Excellence for Nanoscale BioPhotonics, School of Applied Sciences, RMIT University, Melbourne, VIC 3001, Australia

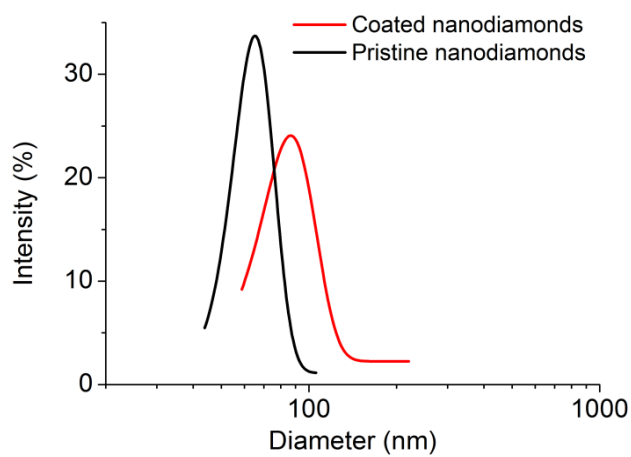

Figure S1. Dynamic light scattering of pristine (black curve) nanodiamonds and nanodiamonds coated with phenol-ionic complexes (red curve)

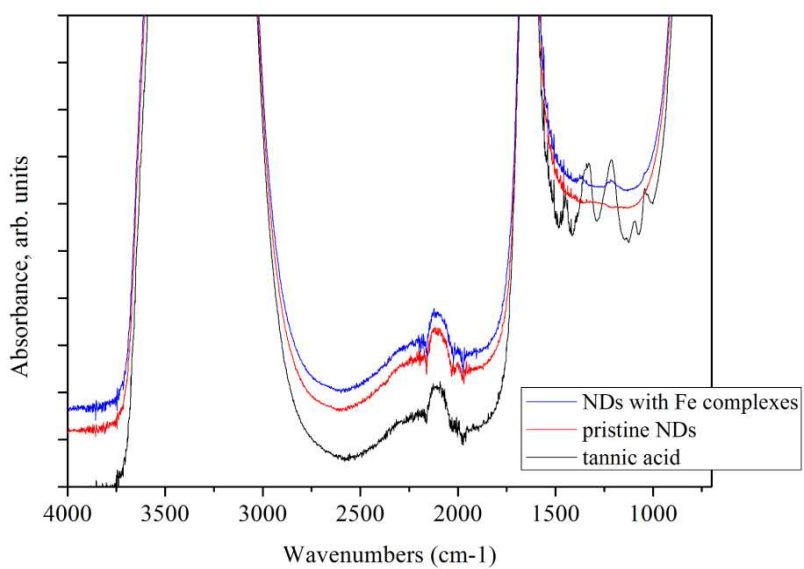

Figure S2. FTIR measurements of the pristine nanodiamonds (red curve), tanic acid (black curve) and NDs with complexes (blue curve).